\documentclass{article}

\usepackage[utf8]{inputenc}
\usepackage[margin=1in]{geometry}
\usepackage[titletoc,title]{appendix}
\usepackage{booktabs}
\usepackage{caption}
\usepackage{subcaption}
\usepackage{amsmath,amsfonts,amssymb,mathtools}

\usepackage{graphicx,float}
\usepackage[affil-it]{authblk}

\usepackage[ruled,vlined]{algorithm2e}
\usepackage{algorithmic}
\usepackage[utf8]{inputenc}

\usepackage{biblatex}
\title{Convolutional neural network for earthquake detection}
\author{José Augusto Proença Maia Devienne}
\affil{Big Data to Earth Scientists}
\date{December, 2020}

\begin{document}

\maketitle

\section{Introduction}

The recent exploitation of natural resources and associated waste water injection in the subsurface have induced many small and moderate earthquakes in the tectonically quiet Central United States. This increase in seismic activity has produced an exponential growth of seismic data recording, which brings the necessity for efficient algorithms to reliably detect earthquakes among this large amount of noisy data. Most current earthquake detection methods are designed for moderate and large events and, consequently, they tend to miss many of the low-magnitude earthquake that are masked by the seismic noise. \textit{Perol et. al} (2018) [1] has focused on the problem of earthquake detection by using a deep-learning approach: the authors proposed a convolutional neural network (ConvNetQuake) to detect and locate earthquake events from seismic records. This reports aims at reproducing part of the methodology  proposed by the author, which is the implementation of a convolutional neural network for classification of events (i.e., earthquake \textit{vs.} noise) from seismic records. 




\section{Methodology}

\begin{figure}[h!]
    \centering
    \includegraphics[width=0.8\linewidth,height=0.8\linewidth]{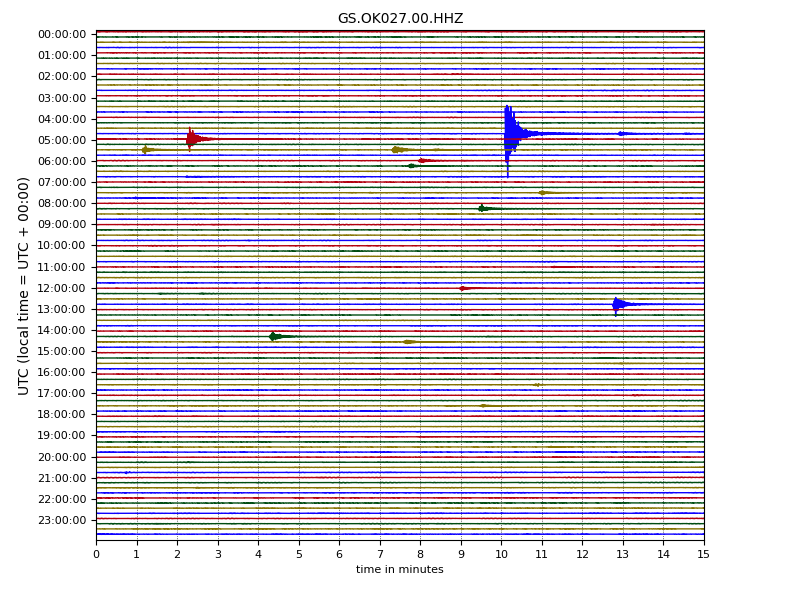}
    \caption{One-day plot (March 1\textsuperscript{st}, 2014) of the horizontal (HHZ) component of the seismic data collected by the station OK027.}
    \label{fig:full_dayl}
\end{figure}

In this report we relied on the seismic data (continuous ground velocity records) collected by the seismic station OK027 and available for download with IRIS API service [2]. The signal is recorded at 100 Hz on three channels, corresponding to the three spatial dimensions: HHZ (up - down), HHN (north - south) and HHE (east - west). The continuous waveform containing a full day (figure \ref{fig:full_dayl}) of seismic recording is initially normalized by subtracting the mean and dividing by the absolute peak amplitude. The normalized record is split into 10-s-long windows and grouped into two categories: windows containing events (i.e., earthquakes) and windows free of events (i.e., containing only seismic noise). The windows containing events was chosen based on the historical catalog of earthquakes of the Ocklahoma Geological Survey (OGS) [3]. Before actually comparing the seismic data with the historical list, the time of earthquake travel was corrected for the travel time of the seismic wave to the station. 

\begin{figure}[h!]
    \centering
    \includegraphics[width=0.9\linewidth,height=0.4\linewidth]{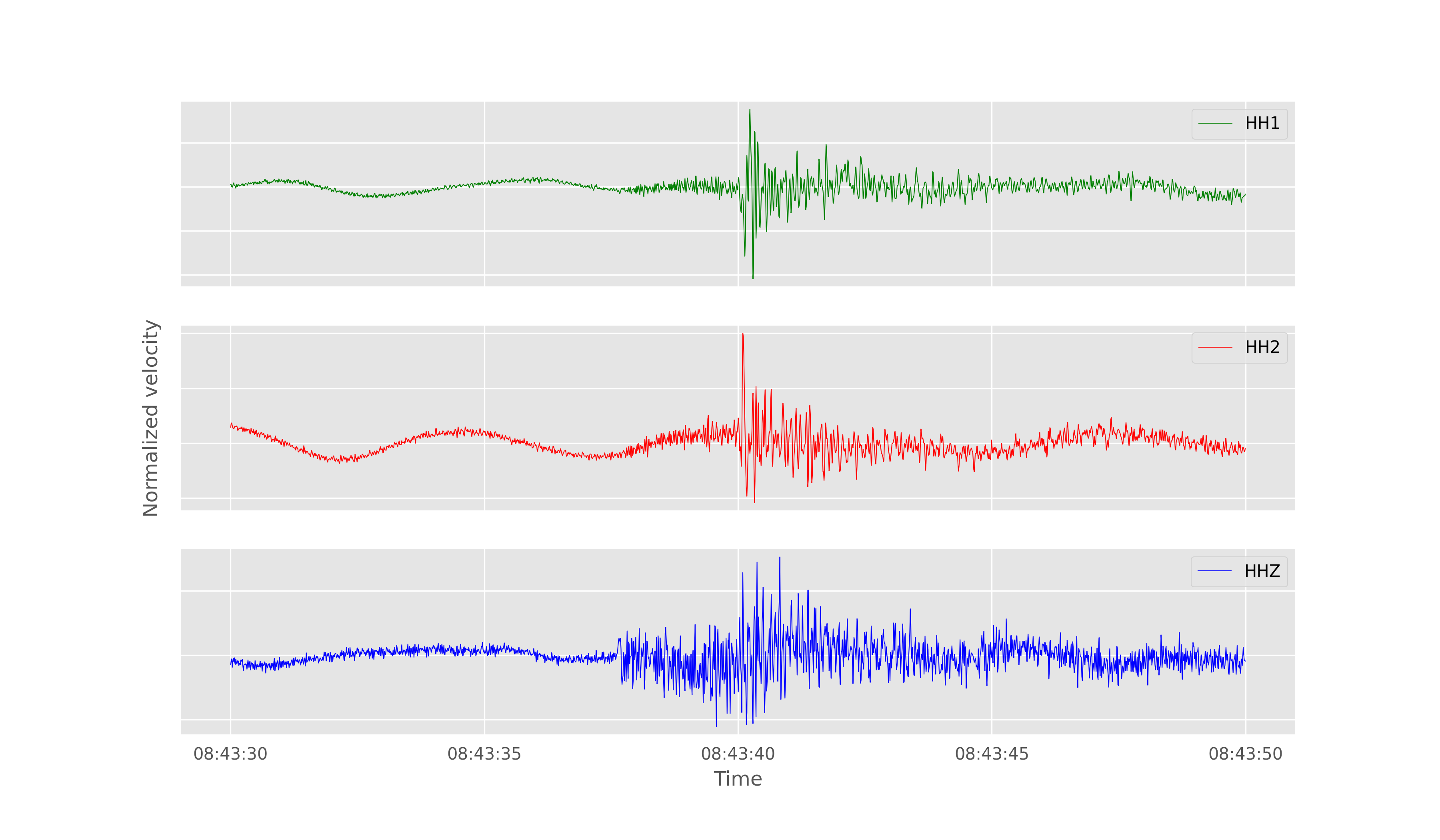}
    \caption{Original seismic data collected in March March 1\textsuperscript{st}, 2014 registering a seismic event at $\sim$ 08:40.}
    \label{fig:orig_data}
\end{figure}

\begin{figure}[h!]
    \centering
    \includegraphics[width=0.9\linewidth,height=0.4\linewidth]{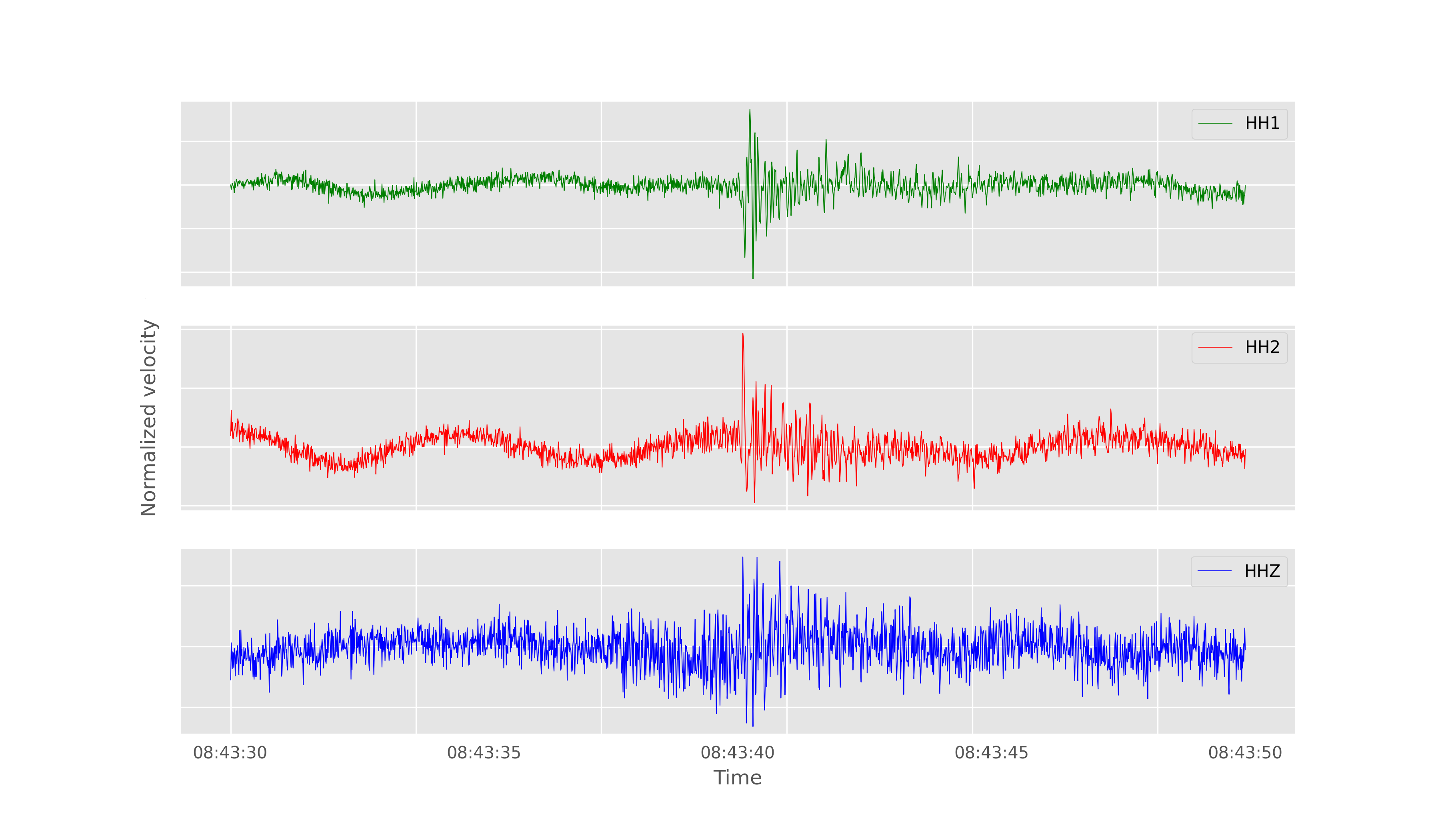}
    \caption{Example of augmented seismic data created by perturbing the original signal (above) with zero-mean Gaussian noise (dispersion ($\sigma$) equal to 2.0 $\times$ 10\textsuperscript{-4} for this case).}
    \label{fig:aug_data}
\end{figure}

The number of 10-sec windows that contain events is much smaller than those that windows containing only seismic noise. Therefore, in order to avoid a overfitting and improve the CNN classification the strategy of data set augmentation was also implemented. This can be done by generating additional event windows from the original ones, perturbing them with zero-mean Gaussian noise (figures \ref{fig:orig_data} and \ref{fig:aug_data}). 

\subsection{Neural Network implementation}

The deep convolutional network model takes a window of three-channel waveform seismogram data as input and predicts its label as either seismic noise or as containing an event. The parameters of the network are optimized to minimize the discrepancy between the predicted and the true label on the training set. The network's input is a two-dimensional tensor Z\textsuperscript{0}\textsubscript{c,t} representing the waveform data for a 10-sec window. The rows of Z\textsuperscript{0}\textsubscript{c,t} for \textit{c} $\in$ \{1, 2, 3\} correspond to the channels of the waveform, whilst \textit{t} is the time index (\textit{t} $\in$ \{1, ... , 1000\}, as 10-sec windows sampled at 100 Hz were considered).

The neural networks was constructed by using eight convolutional layers, followed by one fully connected layer \textit{z} that outputs the classification label. Each channel of the eight convolutional layers was obtained by convolving the channel of the previous layer with a bank of linear filters. The mathematical link between the channels of consecutive layers is made by the activation function (in this report, the activation function used is the rectified linear unit, ReLU). During the training step the network is optimized by minimizing the regularized cross-entropy loss function, which measures the average discrepancy between the predicted and the true value (or class). At the last convolutional layer, the sigmoidal activation function is used, given it's adequacy for binary classification such as the implemented in this report. 

In this report we relied on Keras [4] and Tensoflow [5] to perform the machine learning classification analysis. As mentioned before, the 10-sec windows was previously stored into different directories and classified into either seismic noise or as a windows containing event. These two directories were split again into train and test directories, for training and testing the machine learning predictions. In terms of the number of windows, the test set contains 62.168 windows (29.900 earthquakes and 32.268 noise) and the train set 560.322 contains windows (269.910 earthquakes and 290.412 noise).

\begin{figure}[h!]
    \centering
    \includegraphics[width=0.6\linewidth,height=0.4\linewidth]{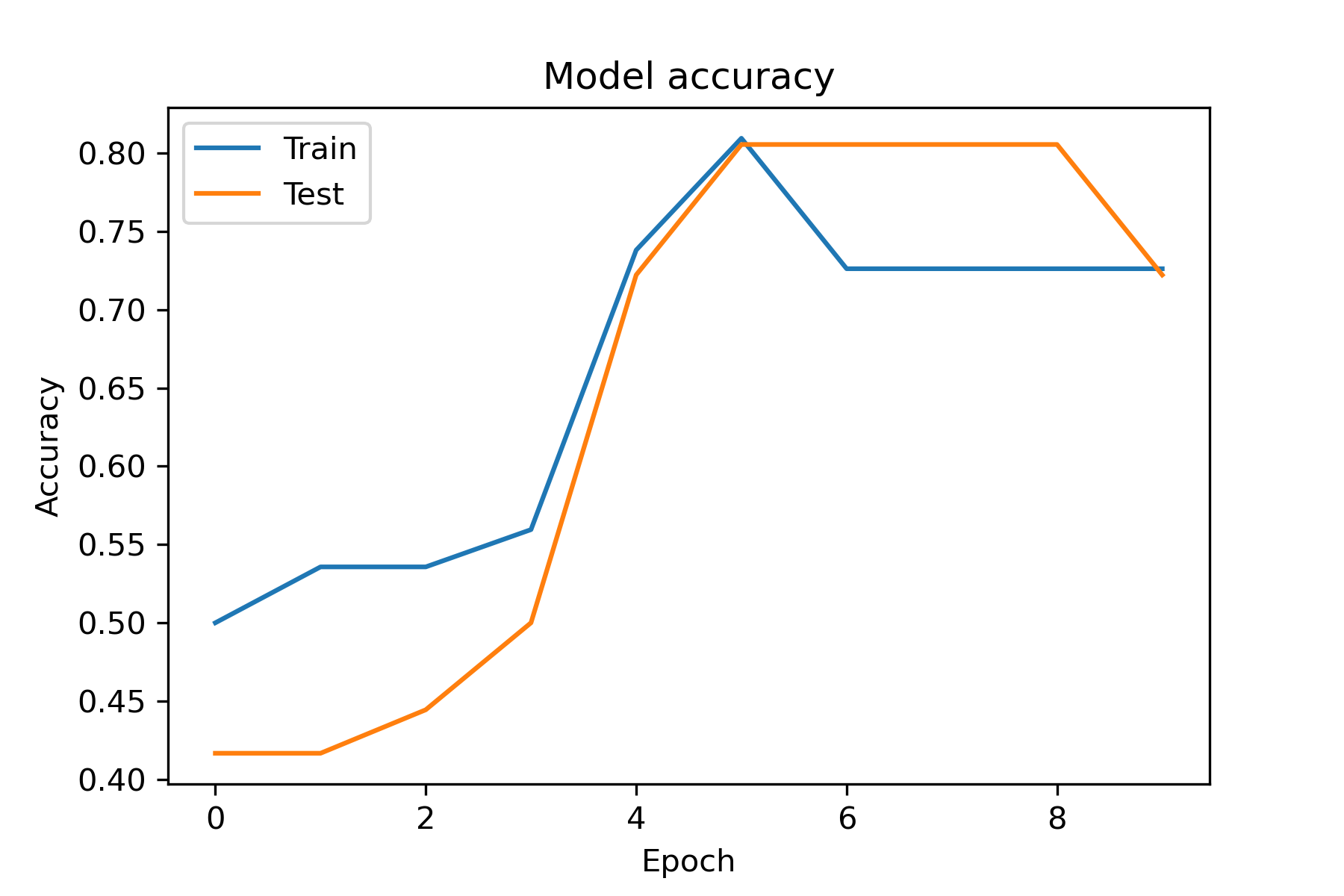}
    \caption{Plot of the accuracy as a function of the epochs.}
    \label{fig:accuracy}
\end{figure}

\begin{figure}[h!]
    \centering
    \includegraphics[width=0.6\linewidth,height=0.4\linewidth]{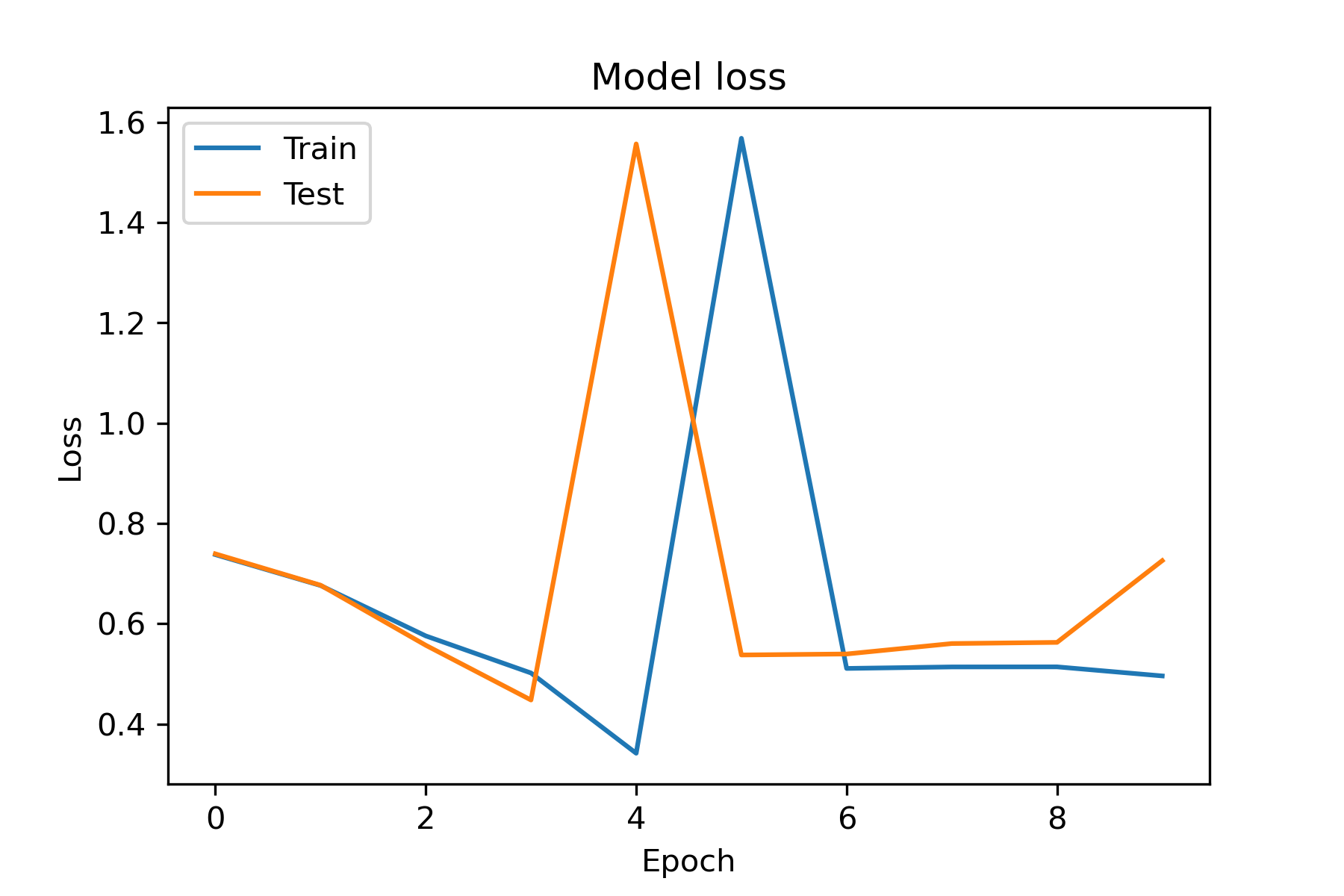}
    \caption{Plot of the cros-entropy loss as a function of the epochs.}
    \label{fig:loss}
\end{figure}

\section{Results and conclusions}

This report aimed at reproducing part of the methodology presented by \textit{Perol et. al} (2018) for the binary classification between seismic noise and earthquake events from the seismic records. The scientific problem addressed by the author considered a more complex analysis, involving not only earthquake detection from seismic noise, but also the probabilistic distribution of the event location. The plots of accuracy and loss function as functions of the epochs make evident that the deep neural network approach to seismic detection represents an accurate method for detecting earthquake from seismic record. The prediction accuracy at the end of the CNN analysis is $\sim$ 72.8\%, which is a fairly good result (in the original paper $\sim$ 94.7\% of accuracy was reached) given the smaller date range considered in this report. Nevertheless, it's also worthy to mention that from the eighth epoch onward, a decrease in the accuracy occurs simultaneously to the growth in the loss function. These behaviours could be more explored considering different numbers of epochs. Due to the computational cost, however, in this report only a run with 10 epochs was investigated. In terms of computational use, the complete run of the CNN used in this report took $\sim$ 4.5 h for training running in a machine with a AMD Ryzen 5 processing unit with 8 GB of RAM memory. An improvement in the training time could be obtained by using a multi worker station, that also can be implemented with Keras and Tensorflow. However, the multi worker stategy was no implemented in this report.  

\vspace*{0.5cm}

\textbf{\Large{References}}

\vspace*{0.5cm}

\noindent [1] Perol et. al, \textit{Sci. Adv.}. 2018.

\noindent [2] IRIS Web Service (\url{http://service.iris.edu/fdsnws/dataselect/1/}).

\noindent [3] Oklahoma Geological Survey (OGS) Web service (\url{ttps://ogsweb.ou.edu/eq_catalog/}). 

\noindent [4] Keras documentation (\url{https://keras.io/getting_started/})

\noindent [5] Tensorflow documentation (\url{https://www.tensorflow.org/tutorials}).

\end{document}